\documentclass[preprint,prd,amsmath,amssymb,amsthm,nofootinbib]{revtex4}
\usepackage[mathscr]{euscript}
\usepackage{bm}
\usepackage{graphicx}
\usepackage{subfigure}
\usepackage[colorlinks=true,linkcolor=red]{hyperref}
\newcommand{\bea}{\begin{eqnarray}}
\newcommand{\eea}{\end{eqnarray}}
\newcommand{\beq}{\begin{equation}}
\newcommand{\eeq}{\end{equation}}

\def\/{\over}

\begin{document}
\title{Spontaneous excitation of an accelerated atom coupled with quantum fluctuations of spacetime}
\author{Shijing Cheng, Jiawei Hu\footnote{Corresponding author: jwhu@hunnu.edu.cn}, Hongwei Yu\footnote{Corresponding author: hwyu@hunnu.edu.cn }}
\affiliation{Department of Physics and Synergetic Innovation Center for Quantum Effects and Applications, Hunan Normal University, Changsha, Hunan 410081, China}

\begin{abstract}

A direct consequence of quantization of gravity would be  quantum gravitational vacuum fluctuations which induce quadrupole moments in gravitationally polarizable atoms. In this paper, we study the spontaneous excitation of a gravitationally polarizable atom with a uniform acceleration $a$ in interaction with a bath of fluctuating quantum gravitational fields in vacuum, and compare the result with that of a static one in a thermal bath of gravitons at the Unruh temperature. We find that, under the fluctuations of spacetime itself, transitions to higher-lying excited  states from the ground state are  possible for both the uniformly accelerated atom in vacuum and the static one in a thermal bath. The appearance of terms in the transition rates  proportional to $a^4$ and $a^2$ indicates that the equivalence between uniform acceleration and thermal field is lost.


\end{abstract}

\maketitle

\section{Introduction}

Based on the classical theory of general relativity, it was predicted by Einstein a hundred years ago that gravitational waves exist as spacetime ripples propagating through the Universe \cite{Einstein}.
The prediction was not directly proved until  signals from black hole merging systems were detected by LIGO \cite{Abbott}. 
Naturally, one may wonder what happens if gravitational waves are quantized. 
One direct consequence when gravity is quantized would be the quantum fluctuations of spacetime itself, which results in the flight time fluctuations of a probe light signal from its source to a detector \cite{Ford95,Yu99,Yu09}. 
Another  effect expected is the Casimir-like force which arises from the quadrupole moments induced by quantum gravitational vacuum fluctuations \cite{Quach15,Holstein,Ford15,Wu16,Hu17,Wu17,Yu18,Pinto16}, in close analogy to the Casimir and the Casimir-Polder forces \cite{Casimir1,Casimir2}. 
Furthermore, quantum fluctuations of spacetime may serve as an environment that provides indirect interactions between the two independent gravitationally polarizable subsystems, which may lead to entanglement generation \cite{Cheng}.

In the present paper, we are concerned with another effect due to the quantum fluctuations of spacetime itself, i.e. the spontaneous emission and excitation of an atom. 
Different physical mechanisms have been put forward to explain why spontaneous emission occurs, such as vacuum fluctuations \cite{Welton,Compagno}, radiation reaction \cite{Ackerhalt}, or a combination of them \cite{Milonni}.
The ambiguity in physical interpretation comes as a result of different choices when ordering commuting operators of the atom and field in a Heisenberg picture approach to the problem. 
It was first suggested by  Dalibard, Dupont-Roc, and Cohen-Tannoudji (DDC)  that when an atom linearly couples to the quantum field, a symmetric operator ordering results in distinctively separable contributions of vacuum fluctuations and radiation reaction to an atomic observable, and furthermore the two contributions are both Hermitian \cite{DDC1,DDC2}. Thus the problem of stability for inertial ground-state atoms in vacuum can be resolved with the DDC prescription \cite{Audretsch}. 
Subsequently, the DDC formalism has been applied to study the radiative properties of an atom in noninertial motion  \cite{Audretsch,Audretsch1,Audretsch2,Passante,Zhu06,Yu06,
Rizzuto07,Rizzuto1,Rizzuto2,Zhu10,Rizzuto11,Zhou16}, in a thermal bath \cite{Tomazelli,Zhu2009}, or in curved spacetime \cite{Zhou072,Zhou10,Zhou12,ZhuZhou,Zhou82}.
When nonlinear atom-field coupling is considered, the mean rate of change of the atomic energy can no longer be separated into  the contributions of vacuum fluctuations and radiation reaction only, and there exists a cross term involving both vacuum fluctuations and radiation reaction which is absent in the linear coupling case, as shown in  Refs.  \cite{Zhou,Li14}.

In this paper, we aim to study the spontaneous excitation of a uniformly accelerated  gravitationally polarizable atom in linear interaction with the fluctuating quantum gravitational fields in vacuum. 
The meaning of a gravitationally polarizable atom is twofold. 
First, it is gravitationally polarizable; i.e. the mass of the atom will be redistributed, and an instantaneous quadrupole moment will be induced under the influence of quantum fluctuations of spacetime itself. 
This is similar to electrically polarizable neutral atoms in which  instantaneous dipoles will be induced by electromagnetic vacuum fluctuations. 
Second, it is quantized and has discrete energy levels. Transitions between the ground state and higher-lying excited states can occur, and a graviton is emitted or absorbed  simultaneously.  
In this paper, we will study the transition rate of a uniformly accelerated gravitationally polarizable atom. In particular, we will  investigate how the result is different from those coupled with matter fields (e.g. scalar and electromagnetic fields), and also compare the result with that of a static atom in a thermal bath of gravitons  at the Unruh temperature. 
Natural units $\hbar=c=32\pi G=1$ will be used in this paper.

\section{The basic formalism} 

We aim to study the  spontaneous excitation of a gravitationally polarizable multilevel atom coupled with a bath of fluctuating quantum gravitational fields.
The atom is assumed to be on a stationary spacetime trajectory $x(\tau)$, with $\tau$ being the proper time of the atom. 
The Hamiltonian describing the time evolution of the atom with respect to the proper time $\tau$ can be written as
\begin{eqnarray}
H_A(\tau)=\sum\limits_{n}\omega_n\sigma_{nn}(\tau),
\end{eqnarray}
where $\sigma_{nn}(\tau)=|n\rangle\langle n|$ and $|n\rangle$ denotes a series of stationary states of the atom with energies $\omega_n$.
The free Hamiltonian of the quantum gravitational field is written as
\begin{eqnarray}
H_F(\tau)=\sum\limits_{k}\omega_{\vec{k}}a^{\dagger}_{\vec{k}}a_{\vec{k}}\frac{dt}{d\tau},
\end{eqnarray}
where $\vec{k}$ denotes the wave vector of the field modes,  $a^{\dagger}_{\vec{k}}$ and $a_{\vec{k}}$ are the creation and annihilation operators with momentum $\vec{k}$, and
$H_I(\tau)$ describes the quadrupolar interaction between the gravitationally polarizable atom and the fluctuating gravitational fields, which can be expressed as
\begin{eqnarray}\label{eq3}
H_I(\tau)=-\frac{1}{2}Q_{ij}(\tau)E_{ij}(x(\tau)),
\end{eqnarray}
where $Q_{ij}(\tau)$ is the induced quadrupole moment operator of the atom, and
$E_{ij}=-\nabla_{i}\nabla_{j} \phi$ with $\phi$ being the gravitational potential.
The quadrupolar interaction Hamiltonian Eq. (\ref{eq3}) can be obtained as follows. 
The energy of a localized mass distribution $\rho_m(x)$ in the presence of an external gravitational potential $\Phi(x)$ is
\begin{equation}\label{v}
V=\int \rho_m(x)\Phi(x)d^3x\;.
\end{equation}
When $\Phi(x)$ varies slowly over the region where the mass is located, it can be expanded as
\begin{equation}\label{phi}
\Phi(x)=\Phi(x_0)+x_i \frac{\partial \Phi(x_0)}{\partial x_i}
    +\frac{1}{2}x_ix_j\frac{\partial^2\Phi(x_0)}{\partial x_i\partial x_j}+\cdots\;,
\end{equation}
so the quadrupolar interaction term reads
\begin{equation}
H_I=\frac{1}{2}\int d^3x \rho_m(x)x_ix_j
   \frac{\partial^2\Phi}{\partial x_i\partial x_j}\;.
\end{equation}
Since $\nabla^2\Phi=0$ in an empty space, the above  equation can be rewritten as
\begin{equation}
H_I=-\frac{1}{2}Q_{ij}E_{ij}\;,
\end{equation}
where
\begin{equation}
Q_{ij}=\int d^3x \rho_m(x)\left(x_ix_j -\frac{1}{3}\delta_{ij}r^2 \right)\;
\end{equation}
and
\begin{equation}\label{E_ij}
E_{ij}=-\frac{\partial^2\Phi}{\partial x_i\partial x_j}
       +\frac{1}{3}\delta_{ij}\nabla^2\Phi\;.
\end{equation}
In general relativity,  $E_{ij}$ is defined as the Weyl tensor $C_{i0j0}$,  which coincides with  Eq. (\ref{E_ij}) in the Newtonian limit. Here $E_{ij}=C_{i0j0}$ and its dual tensor $B_{ij}=-\frac{1}{2}\epsilon_{ikl}{C^{kl}}_{j0}$ are the gravitoelectric and gravitomagnetic tensors which satisfy the linearized Einstein field equations organized in a form similar to the Maxwell equations~\cite{Campbell76,matte53,campbell71,Szekeres,maartens98,Ruggiero02,ramos10,Ingraham}.

With the total Hamiltonian  $H=H_A(\tau)+H_F(\tau)+H_I(\tau)$, one obtains the Heisenberg equations of motion for the dynamical variables of the atom and the gravitational field as
\begin{eqnarray}
\frac{d}{d\tau}\sigma_{mn}(\tau)&=&i(\omega_m-\omega_n)\sigma_{mn}(\tau)-\frac{i}{2}E_{ij}\left(x(\tau)\right)\left[Q_{ij}(\tau),\sigma_{mn}(\tau)\right],\\ \nonumber
\frac{d}{dt}a_{\vec{k}}(t)&=&-i\omega_{\vec{k}}a_{\vec{k}}(t)-\frac{i}{2}Q_{ij}(\tau)\left[E_{ij}(x(\tau)),a_{\vec{k}}(t(\tau))\right]\frac{d\tau}{dt}.
\end{eqnarray}
Solving the equations above and separating the ``free'' and ``source" parts of the dynamical variables, we have
\begin{eqnarray}
\sigma_{mn}(\tau)=\sigma_{mn}^F(\tau)+\sigma_{mn}^S(\tau),\ \ \ a_{\vec{k}}(t)=a_{\vec{k}}^F(t)+a_{\vec{k}}^S(t),
\end{eqnarray}
where
\begin{eqnarray}
\sigma_{mn}^F(\tau)&=&\sigma_{mn}^F(\tau_0)e^{i(\omega_m-\omega_n)(\tau-\tau_0)},\nonumber\\ \nonumber
\sigma_{mn}^S(\tau)&=&-\frac{i}{2}\int_{\tau_0}^{\tau}d\tau' E_{ij}^F(x(\tau'))\left[Q_{ij}^F(\tau'),\sigma_{mn}^F(\tau)\right],\\ \nonumber
a_{\vec{k}}^F(t(\tau))&=&a_{\vec{k}}^F(t(\tau_0))e^{-i\omega_{\vec{k}}(t(\tau)-t(\tau_0))},\\ 
a_{\vec{k}}^S(t(\tau))&=&-\frac{i}{2}\int_{\tau_0}^{\tau}d\tau' Q_{ij}^F(\tau')\left[E_{ij}^F(x(\tau')),a_{\vec{k}}^F(t(\tau))\right].
\end{eqnarray}
With the symmetric ordering \cite{DDC1,DDC2}, the equation of motion for the  energy $H_A(\tau)$ in the interaction representation can be separated into two parts, i.e. the vacuum fluctuations (VF) and the radiation reaction (RR),
\begin{eqnarray}
\left(\frac{d}{d\tau}H_{A}(\tau)\right)=\left(\frac{d}{d\tau}H_{A}(\tau)\right)_{VF}+\left(\frac{d}{d\tau}H_{A}(\tau)\right)_{RR},
\end{eqnarray}
where
\begin{eqnarray}
\left(\frac{d}{d\tau}H_{A}(\tau)\right)_{VF}&=&-\frac{i}{4}\biggl\{E_{ij}^F(x(\tau)),\left[Q_{ij}(\tau),\sum\limits_n\omega_n\sigma_{nn} (\tau)\right]\biggl\},\nonumber\\
\left(\frac{d}{d\tau}H_{A}(\tau)\right)_{RR}&=&-\frac{i}{4}\biggl\{E_{ij}^S(x(\tau)),\left[Q_{ij}(\tau),\sum\limits_n\omega_n\sigma_{nn} (\tau)\right]\biggl\}.
\end{eqnarray}
We assume that initially the field is  in state $|a\rangle$ (vacuum or thermal state), while the atom is in state $|b\rangle$. Taking the expectation values of $\left(\frac{dH_{A}(\tau)}{d\tau}\right)_{VF(RR)}$, we have
\begin{eqnarray}\label{evolution1}
\left\langle \frac{d}{d\tau}H_{A}(\tau)\right\rangle_{VF}&=&\frac{i}{2}\int_{\tau_0}^{\tau}d\tau'C_{ijkl}^F(x(\tau),x(\tau')) \frac{d}{d\tau}(\chi_{ijkl}^A)_b(\tau,\tau'),
\end{eqnarray}
\begin{eqnarray}\label{evolution2}
\left\langle \frac{d}{d\tau}H_{A}(\tau)\right\rangle_{RR}&=&\frac{i}{2}\int_{\tau_0}^{\tau}d\tau'\chi_{ijkl}^F(x(\tau),x(\tau')) \frac{d}{d\tau}(C_{ijkl}^A)_b(\tau,\tau'),
\end{eqnarray}
where
$|\rangle$=$|a,b\rangle$.
Here, the statistical functions $C_{ijkl}^F$ and $\chi_{ijkl}^F$ are the symmetric correlation function and linear susceptibility of the gravitational field respectively, defined as
\begin{eqnarray}\label{fieldC}
C_{ijkl}^F(x(\tau),x(\tau'))&=&\frac{1}{2}\langle a\left|\left\{E^F_{ij}(x(\tau)),E^F_{kl}(x(\tau'))\right\}\right|a\rangle,
\end{eqnarray}
\begin{eqnarray}\label{fieldX}
\chi_{ijkl}^F(x(\tau),x(\tau'))&=&\frac{1}{2}\langle a\left|\left[E^F_{ij}(x(\tau)),E^F_{kl}(x(\tau'))\right]\right|a\rangle,
\end{eqnarray}
and
\begin{eqnarray}
(C_{ijkl}^A)_b(\tau,\tau')&=&\frac{1}{2}\langle b\left|\left\{Q_{ij}^F(\tau),Q_{kl}^F(\tau')\right\}\right|b\rangle,\nonumber\\
(\chi_{ijkl}^A)_b(\tau,\tau')&=&\frac{1}{2}\langle b\left|\left[Q_{ij}^F(\tau),Q_{kl}^F(\tau')\right]\right|b\rangle
\end{eqnarray}
are the symmetric correlation function and the linear susceptibility of the atom.
It is obvious that $(\chi_{ijkl}^A)_b$ and $(C_{ijkl}^A)_b$ do not rely on the trajectory of the atom, and their  explicit forms can be given  as follows:
\begin{eqnarray}
(C_{ijkl}^A)_b(\tau,\tau')&=&\frac{1}{2}\sum\limits_{\omega_{bd}}\biggl[\langle b|Q_{ij}^F(0)|d\rangle \langle d|Q_{kl}^F(0)|b\rangle e^{i\omega_{bd}(\tau-\tau')}\nonumber\\
&&\ \ \ \ \ \ +\langle b|Q_{kl}^F(0)|d\rangle \langle d|Q_{ij}^F(0)|b\rangle e^{-i\omega_{bd}(\tau-\tau')}\biggl],\nonumber\\
(\chi_{ijkl}^A)_b(\tau,\tau')&=&
\frac{1}{2}\sum\limits_{\omega_{bd}}\biggl[\langle b|Q_{ij}^F(0)|d\rangle \langle d|Q_{kl}^F(0)|b\rangle e^{i\omega_{bd}(\tau-\tau')}\nonumber\\
&&\ \ \ \ \ \ -\langle b|Q_{kl}^F(0)|d\rangle \langle d|Q_{ij}^F(0)|b\rangle e^{-i\omega_{bd}(\tau-\tau')}\biggl].
\end{eqnarray}
Here $\omega_{bd}=\omega_{b}-\omega_{d}$, and the sum extends over a complete set of states of the atom.

\section{Spontaneous excitation of a uniformly accelerated gravitationally polarizable atom}

In this section, we study the spontaneous excitation of a gravitationally polarizable multilevel atom moving with a constant  proper acceleration in vacuum. 
We assume that the atom accelerates along the $x$ direction, so the trajectory can be  written as
\begin{eqnarray}
t(\tau)=\frac{1}{a}\sinh{a\tau},\ \  x(\tau)=\frac{1}{a}\cosh{a\tau},\ \ y(\tau)=z(\tau)=0,
\end{eqnarray}
where $\tau$ is the proper time, and $a$ is the proper acceleration. 

The spacetime metric $g_{\mu\nu}$ can be expressed as a sum of the flat spacetime metric $\eta_{\mu\nu}$ and a linearized perturbation $h_{\mu\nu}$. 
In the transverse traceless  gauge, the spacetime perturbation can be quantized as  \cite{Yu99}
\begin{equation}
h_{ij}=\sum_{\mathbf{k},\lambda}[a_{\mathbf{k},\lambda}e_{ij}(\mathbf{k},\lambda)f_{\mathbf{k}}+{\rm H.c.}],
\end{equation}
where $f_{\mathbf{k}}=(2\omega(2\pi)^{3})^{-\frac{1}{2}}e^{i(\mathbf{k}\cdot\mathbf{x}-\omega t)}$ is the field mode, and $e_{\mu\nu}(\mathbf{k},\lambda)$ is the  polarization tensor with $\omega=|\mathbf{k}|=(k_{x}^{2}+k_{y}^{2}+k_{z}^{2})^{\frac{1}{2}}$.
Here H.c. denotes the Hermitian conjugate, and $\lambda$ labels the polarization state.
From the definition of $E_{ij}$ ($E_{ij}=C_{i0j0}$), we have
\begin{equation}\label{E}
E_{ij}=\frac{1}{2}\ddot{h}_{ij},
\end{equation}
where a dot means $\frac{\partial}{\partial t}$.
Then the two point function for the gravitational field in the vacuum state $|0\rangle$  in the laboratory frame can be obtained as~\cite{Yu99}
\begin{equation}\label{correlation1}
{\langle{0|E_{ij}(x)E_{kl}(x')|0}\rangle}=\frac{1}{8(2\pi)^{3}}\int d^{3}\mathbf{k}\sum_{\lambda}e_{ij}(\mathbf{k},\lambda)e_{kl}(\mathbf{k},\lambda)\,{\omega^{3}}e^{i\mathbf{k}\cdot(\mathbf{x}-\mathbf{x'})}e^{-i\omega(t-t')},
\end{equation}
where
\bea\label{correlation2}
&&\sum_{\lambda}e_{ij}(\mathbf{k},\lambda)e_{kl}(\mathbf{k},\lambda)=\delta_{ik}\delta_{jl}+\delta_{il}\delta_{jk}-\delta_{ij}\delta_{kl}+\hat{k_{i}}\hat{k_{j}}\hat{k_{k}}\hat{k_{l}}+\hat{k_{i}}\hat{k_{j}}\delta_{kl}+\hat{k_{k}}\hat{k_{l}}\delta_{ij}\nonumber\\
&&\qquad\qquad\qquad\qquad\qquad\qquad\qquad\qquad-\hat{k_{i}}\hat{k_{l}}\delta_{jk}-\hat{k_{i}}\hat{k_{k}}\delta_{jl}-\hat{k_{j}}\hat{k_{l}}\delta_{ik}-\hat{k_{j}}\hat{k_{k}}\delta_{il},
\eea
with $\hat{k_{i}}={k_{i}}/{k}$.
The symmetric correlation function $C_{ijkl}^F$ and the linear susceptibility $\chi_{ijkl}^F$ according to Eqs. (\ref{fieldC}) and (\ref{fieldX}), with a Lorentz
transformation from the laboratory frame to the frame of the atom, can be calculated as
\begin{eqnarray}\nonumber
C_{1111}^F(x(\tau),x(\tau'))&=&-\frac{a^6}{32\pi^2} \Delta^{+}_{\mathrm{vac.}},\ \ \ \  \chi_{1111}^F(x(\tau),x(\tau'))=-\frac{a^6}{32\pi^2} \Delta^{-}_{\mathrm{vac.}},\\\nonumber
C_{1122}^F(x(\tau),x(\tau'))&=&\frac{a^6}{64\pi^2}\Delta^{+}_{\mathrm{vac.}},\ \ \ \ \ \  \chi_{1122}^F(x(\tau),x(\tau'))=\frac{a^6}{64\pi^2}\Delta^{-}_{\mathrm{vac.}},\\\nonumber
C_{1212}^F(x(\tau),x(\tau'))&=&-\frac{3a^6}{128\pi^2} \Delta^{+}_{\mathrm{vac.}},  \ \ \chi_{1212}^F(x(\tau),x(\tau'))=-\frac{3a^6}{128\pi^2} \Delta^{-}_{\mathrm{vac.}},\\
&&
\end{eqnarray}
with
\bea\label{sinh}
\Delta^{+}_{\mathrm{vac.}}&=&\sinh^{-6}{\left[\frac{a(\tau-\tau'-i\epsilon)}{2}\right]}+\sinh^{-6}{\left[\frac{a(\tau-\tau'+i\epsilon)}{2}\right]},\nonumber \\
\Delta^{-}_{\mathrm{vac.}}&=&\sinh^{-6}{\left[\frac{a(\tau-\tau'-i\epsilon)}{2}\right]}-\sinh^{-6}{\left[\frac{a(\tau-\tau'+i\epsilon)}{2}\right]}.
\eea
Here the nonzero components of $C_{ijkl}^F$ and $\chi_{ijkl}^F$ satisfy the following relations,
\bea\nonumber
X_{1111}^F&=&X_{2222}^F=X_{3333}^F,\\\nonumber 
X_{1122}^F&=&X_{2211}^F=X_{1133}^F=X_{3311}^F=X_{2233}^F=X_{3322}^F,\\ \nonumber
X_{1212}^F&=&X_{1221}^F=X_{2112}^F=X_{2121}^F=X_{1313}^F=X_{1331}^F\\
&=&X_{3113}^F=X_{3131}^F=X_{2323}^F=X_{2332}^F=X_{3223}^F=X_{3232}^F,
\eea
where $X_{ijkl}^F$ denotes $C_{ijkl}^F$ or $\chi_{ijkl}^F$.
With a substitution $u=\tau-\tau'$, and an extension of the range of integration to infinity for sufficiently long times $\tau-\tau_0$ in Eqs. (\ref{evolution1}) and (\ref{evolution2}), the contributions of vacuum fluctuations and radiation reaction to the average rate of change of the atomic energy can be obtained with some straightforward calculations as
\begin{eqnarray}\label{vf}
\left\langle \frac{d}{d\tau}H_{A}(\tau)\right\rangle_{VF}&=&-\frac{1}{4}\sum\limits_{\omega_{bd}}\omega_{bd}\biggl[|\langle b|Q_{11}^F(0)|d\rangle|^2+|\langle b|Q_{22}^F(0)|d\rangle|^2+|\langle b|Q_{33}^F(0)|d\rangle|^2\biggl] \mathcal{G}^F_{1111}\nonumber\\
&&-\frac{1}{4}\sum\limits_{\omega_{bd}}\omega_{bd}\biggl[\langle b|Q_{11}^F(0)|d\rangle \langle d|Q_{22}^F(0)|b\rangle+\langle b|Q_{22}^F(0)|d\rangle \langle d|Q_{11}^F(0)|b\rangle \nonumber\\ \nonumber
&&\qquad\quad\quad\ \ +\langle b|Q_{11}^F(0)|d\rangle \langle d|Q_{33}^F(0)|b\rangle+\langle b|Q_{33}^F(0)|d\rangle \langle d|Q_{11}^F(0)|b\rangle\\ \nonumber
&&\qquad\quad\quad\ \ +\langle b|Q_{22}^F(0)|d\rangle \langle d|Q_{33}^F(0)|b\rangle+\langle b|Q_{33}^F(0)|d\rangle \langle d|Q_{22}^F(0)|b\rangle\biggl] \mathcal{G}^F_{1122}\\ 
&&-\sum\limits_{\omega_{bd}}\omega_{bd}\biggl[|\langle b|Q_{12}^F(0)|d\rangle|^2+|\langle b|Q_{13}^F(0)|d\rangle|^2+|\langle b|Q_{23}^F(0)|d\rangle|^2\biggl]\mathcal{G}^F_{1212}
\end{eqnarray}
and
\begin{eqnarray}\label{rr}
\left\langle \frac{d}{d\tau}H_{A}(\tau)\right\rangle_{RR}&=&-\frac{1}{4}\sum\limits_{\omega_{bd}}\omega_{bd}\biggl[|\langle b|Q_{11}^F(0)|d\rangle|^2+|\langle b|Q_{22}^F(0)|d\rangle|^2+|\langle b|Q_{33}^F(0)|d\rangle|^2\biggl] \mathcal{K}^F_{1111}\nonumber\\ \nonumber
&&-\frac{1}{4}\sum\limits_{\omega_{bd}}\omega_{bd}\biggl[\langle b|Q_{11}^F(0)|d\rangle \langle d|Q_{22}^F(0)|b\rangle+\langle b|Q_{22}^F(0)|d\rangle \langle d|Q_{11}^F(0)|b\rangle\\ \nonumber
&&\qquad\quad\quad\ \ +\langle b|Q_{11}^F(0)|d\rangle \langle d|Q_{33}^F(0)|b\rangle+\langle b|Q_{33}^F(0)|d\rangle \langle d|Q_{11}^F(0)|b\rangle\\ \nonumber
&&\qquad\quad\quad\ \ +\langle b|Q_{22}^F(0)|d\rangle \langle d|Q_{33}^F(0)|b\rangle+\langle b|Q_{33}^F(0)|d\rangle \langle d|Q_{22}^F(0)|b\rangle\biggl] \mathcal{K}^F_{1122}\\ 
&&-\sum\limits_{\omega_{bd}}\omega_{bd}\biggl[|\langle b|Q_{12}^F(0)|d\rangle|^2+|\langle b|Q_{13}^F(0)|d\rangle|^2+|\langle b|Q_{23}^F(0)|d\rangle|^2\biggl] \mathcal{K}^F_{1212},
\end{eqnarray}
where
\begin{eqnarray}
\mathcal{G}^F_{ijkl}=\int_{-\infty}^{\infty}du\  e^{i\omega_{bd}u}C_{ijkl}^F(u) ,\ \ \ \mathcal{K}^F_{ijkl}=\int_{-\infty}^{\infty}du\  e^{i\omega_{bd}u}\chi_{ijkl}^F(u) 
\end{eqnarray}
are the Fourier transforms of $C_{ijkl}^F$ and $\chi_{ijkl}^F$.

For a concrete example, we assume $\langle b|Q_{11}^F(0)|d\rangle=-\langle b|Q_{22}^F(0)|d\rangle=Q$ and other components are zero, in accordance with the requirement that the  quadrupole operator is symmetric and traceless.
The contributions of vacuum fluctuations and radiation reaction to the mean rate of change of the  energy are respectively 
\begin{eqnarray}\label{vac1}\nonumber
\left\langle \frac{d}{d\tau}H_{A}(\tau)\right\rangle_{VF}&=&-\frac{Q^2}{40\pi}\sum\limits_{\omega_{bd}>0}\omega_{bd}^6 \left(1+\frac{2}{e^{2\pi \omega_{bd}/a}-1}\right)\left(1+\frac{5a^2}{\omega_{bd}^2}+\frac{4a^4}{\omega_{bd}^4}\right) \\ 
&&+\frac{Q^2}{40\pi}\sum\limits_{\omega_{bd}<0}\omega_{bd}^6 \left(1+\frac{2}{e^{2\pi |\omega_{bd}|/a}-1}\right)\left(1+\frac{5a^2}{\omega_{bd}^2}+\frac{4a^4}{\omega_{bd}^4}\right)
\end{eqnarray}
and
\begin{eqnarray}\label{the1}
\left\langle \frac{d}{d\tau}H_{A}(\tau)\right\rangle_{RR}&=&-\frac{Q^2}{40\pi}\sum\limits_{\omega_{bd}}\omega_{bd}^6 \left(1+\frac{5a^2}{\omega_{bd}^2}+\frac{4a^4}{\omega_{bd}^4}\right).
\end{eqnarray}
This shows that vacuum fluctuations lead to not only  excitation of an accelerated ground-state atom, but also deexcitation of an excited-state one equally, while radiation reaction always diminishes the atomic energy no matter if the atom is initially in the ground state or higher-lying excited states, just as that of a uniformly accelerated atom linearly coupled to vacuum scalar \cite{Audretsch} or electromagnetic fields \cite{Zhu06,Yu06}, or nonlinearly coupled to vacuum Dirac \cite{Zhou} or Rarita-Schwinger fields  \cite{Li14}. 
Note that in the nonlinear coupling case \cite{Zhou,Li14}, it is the cross term involving both vacuum fluctuations and radiation reaction that plays the role of radiation reaction  in the linear coupling case.
The total rate of change of the atomic energy (TOT) for accelerated (acc) atoms is
\begin{eqnarray}\label{vac}\nonumber
\left\langle \frac{d}{d\tau}H_{A}(\tau)\right\rangle_{\mathrm{TOT,acc}}&=&-\frac{Q^2}{20\pi}\sum\limits_{\omega_{bd}>0}\omega_{bd}^6 \left(1+\frac{1}{e^{2\pi \omega_{bd}/a}-1}\right) \left(1+\frac{5a^2}{\omega_{bd}^2}+\frac{4a^4}{\omega_{bd}^4}\right) \\ 
&&+\frac{Q^2}{20\pi}\sum\limits_{\omega_{bd}<0}\frac{\omega_{bd}^6}{e^{2\pi |\omega_{bd}|/a}-1} \left(1+\frac{5a^2}{\omega_{bd}^2}+\frac{4a^4}{\omega_{bd}^4}\right).
\end{eqnarray}
It is obvious that the transition to the higher-lying states of an  accelerated ground-state atom is allowed in  vacuum.

An observer with a uniform acceleration perceives the Minkowski vacuum as a thermal bath at a temperature proportional to its acceleration, which is known as the Unruh effect~\cite{Unruh}. 
In the following, we will compare the result above with that for a static atom immersed in a thermal bath of gravitons. 
The corresponding two point function of gravitational fields  takes the form
\begin{eqnarray}
&&\langle \beta|{E_{ij}(x(\tau)) E_{kl}(x(\tau'))}|\beta\rangle\nonumber \\
&=&\frac{1}{8(2\pi)^{3}} \sum\limits_{m=-\infty}^{\infty}\int d^{3}\mathbf{k}\sum\limits_{\lambda}e_{ij}(\mathbf{k},\lambda)e_{kl}(\mathbf{k},\lambda)\ {\omega^{3}}e^{-i\omega(\tau-\tau'-im\beta)},
\end{eqnarray}
where 
$\beta=1/(kT)$.
With the same assumption of $\langle b|Q_{11}^F(0)|d\rangle=-\langle b|Q_{22}^F(0)|d\rangle=Q$, the total rate of change of the excitation energy of the atom immersed in the thermal bath (tb) is
\bea\label{the}
\nonumber
\left\langle \frac{d}{d\tau}H_{A}(\tau)\right\rangle_{\mathrm{TOT,tb}}=&&-\frac{Q^2}{20\pi}\sum\limits_{\omega_{bd}>0}\omega_{bd}^6 \left(1+\frac{1}{e^{\beta \omega_{bd}}-1}\right) \nonumber\\&&+\frac{Q^2}{20\pi}\sum\limits_{\omega_{bd}<0} \frac{\omega_{bd}^6}{e^{\beta |\omega_{bd}|}-1}.
\eea
So, the transition to the higher-lying states is possible. 

A comparison between Eqs. (\ref{vac}) and (\ref{the}) shows that the transition rates of a uniformly accelerated atom coupled with gravitational vacuum fluctuations are not exactly the same as that of a static atom in a thermal bath, due to the appearance of two terms proportional to $a^4$ and $a^2$. The $a^4$ and $a^2$ terms also exist in the Dirac field case, while the coupling between the atom and the Dirac field is nonlinear \cite{Zhou}.  When $a/\omega_{bd}\gg 1$, terms proportional to $a^4$ and $a^2$ become dominant. Therefore, the equivalence between uniform acceleration and thermal field is lost.
Similar conclusions have been drawn in the electromagnetic field and Rarita-Schwinger field cases, while the nonthermal term  is proportional to $a^2$ in the electromagnetic field case  \cite{Passante,Yu06}, and up to $a^8$ in the Rarita-Schwinger field case \cite{Li14}. 
In fact, the effect of vacuum fluctuations on the rate of change of the atomic energy for a uniformly accelerated atom is fully equivalent to that of a thermal field only when an atom is in interaction with the fluctuating scalar fields in the free Minkowski vacuum  \cite{Audretsch}.
Nevertheless, the asymptotic equilibrium state of uniformly accelerated atoms, which can be derived from the transition rates Eq. (\ref{vac}), is exactly a thermal state at the Unruh temperature, although reached in a different way compared with the static atoms in a thermal bath.

The appearance of power terms in acceleration in the transition rates is a result of the derivative coupling nature of the interaction. The electric field strength $E_{i}$ can be expressed as the derivative of the electromagnetic vector potential, and the gravitoelectric field $E_{ij}$ can be expressed as the second order derivative of the metric tensor (gravitational potential). The derivatives of the Wightman function increase the order of the pole in the $\sinh$ function [in e.g. Eq~(\ref{sinh})], and the higher the order of the pole, the higher the powers of $a$ in the transition rates. Therefore, there exist extra $a^2$ terms in the electromagnetic field case, and extra $a^4$ and  $a^2$ terms in the gravitational field case compared with the case when the atom is coupled to the scalar field via  monopole coupling.
For the same reason, when nonlinear atom-field coupling is considered, there exist terms proportional to $a^6$ and $a^8$ in the Rarita-Schwinger field case~\cite{Li14} compared to the Dirac field case \cite{Zhou}. Actually, power terms in $a$ will also appear in the transition rates in the scalar field case when the monopole coupling is replaced by  a derivative coupling \cite{a1,a2,a3}.

Now an important question is how large is the effect? To obtain some numerical estimations, we rewrite  the total rate of change of the atomic energy (\ref{vac}) in the International System of Units as
\begin{eqnarray}\label{vac1}\nonumber
\left\langle \frac{d}{d\tau}H_{A}(\tau)\right\rangle_{\mathrm{TOT,acc}}&=&-\frac{8 G Q^2}{5 c^5}\sum\limits_{\omega_{bd}>0}\omega_{bd}^6 \left(1+\frac{1}{e^{2\pi c \omega_{bd}/a}-1}\right) \left(1+\frac{5a^2}{c^2\omega_{bd}^2}+\frac{4a^4}{c^4\omega_{bd}^4}\right) \\ 
&&+\frac{8 G Q^2}{5 c^5}\sum\limits_{\omega_{bd}<0}\frac{\omega_{bd}^6}{e^{2\pi c |\omega_{bd}|/a}-1} \left(1+\frac{5a^2}{c^2 \omega_{bd}^2}+\frac{4a^4}{c^4 \omega_{bd}^4}\right).
\end{eqnarray}  
In analogy to electrodynamics, we define a gravitational polarizability $\alpha\equiv \frac{Q^2}{\hbar \omega}$, which can be derived from the geodesic deviation equation, and is found to be $\alpha\sim \frac{M R^2}{\omega^2}$ \cite{Ford15}, where  $M$, $R$, and $\omega$ are the mass, radius, and the frequency respectively.  
Now, we assume that the gravitationally polarizable atom is composed of two point masses $M_1$ and $M_2$, which are bounded by gravity. 
In analogy to the hydrogen atom, such a gravitationally bound system also has discrete energy levels $E_n=-\frac{G^2 {M_1}^3 {M_2}^3}{2 \hbar ^2 n^2 (M_1+M_2)}$, and Bohr radius $R=\frac{ \hbar ^2 (M_1+M_2)}{G {M_1}^2 {M_2}^2}$. 
For such an inertial atom in vacuum, the emission rate 
can be calculated as 
$\Gamma_{0\downarrow}\equiv\frac{\langle\frac{d}{d\tau}H_{A}(\tau)\rangle_{\mathrm{TOT}}}{\hbar \omega}=\frac{8 G (M_1+M_2) R^2 {\omega}^4} {5 c^5}=1.86\times 10^{42} \frac{M_1^8M_2^8}{(M_1+M_2)m_{Pl}^{15}}s^{-1}$, where $m_{Pl}=\sqrt{\hbar c/G}$ is the Planck mass. 
It is obvious that the transition rate increases with the mass of the atom. However, the mass cannot be arbitrarily large, since it is related to the radius of the atom. One 
expects that the Bohr radius should be larger than the corresponding Schwarzschild radius ${2GM}/{c^2}$, as well as larger than the Planck length $l_{Pl}=\sqrt{\hbar G/c^3}$, both of which require that $M$ should be smaller than the Planck mass $m_{Pl}$.
On the other hand, if we want to observe such transitions, the lifetime of the excited state $\Gamma_{0\downarrow}^{-1}$ should be at least smaller than the  age of the Universe, which requires that the mass should be larger than $10^{-4}m_{Pl}$. For simplicity,  we assume that $M_1=M_2=M$, then the transition rate for a gravitationally polarizable inertial atom in vacuum $\Gamma_{0\downarrow}$ ranges from $9.3\times 10^{-19}~s^{-1}$ ($M\sim10^{-4}m_{Pl}$) to $9.3\times 10^{41}~s^{-1}$ ($M\sim m_{Pl}$). For reference, the corresponding Bohr radius lies in $3.2\times 10^{-35}~m<R<3.2\times 10^{-23}~m$, which is much smaller than the radius of an atomic nucleus.



\begin{figure}[!htbp]
\centering\includegraphics[width=0.5\textwidth]{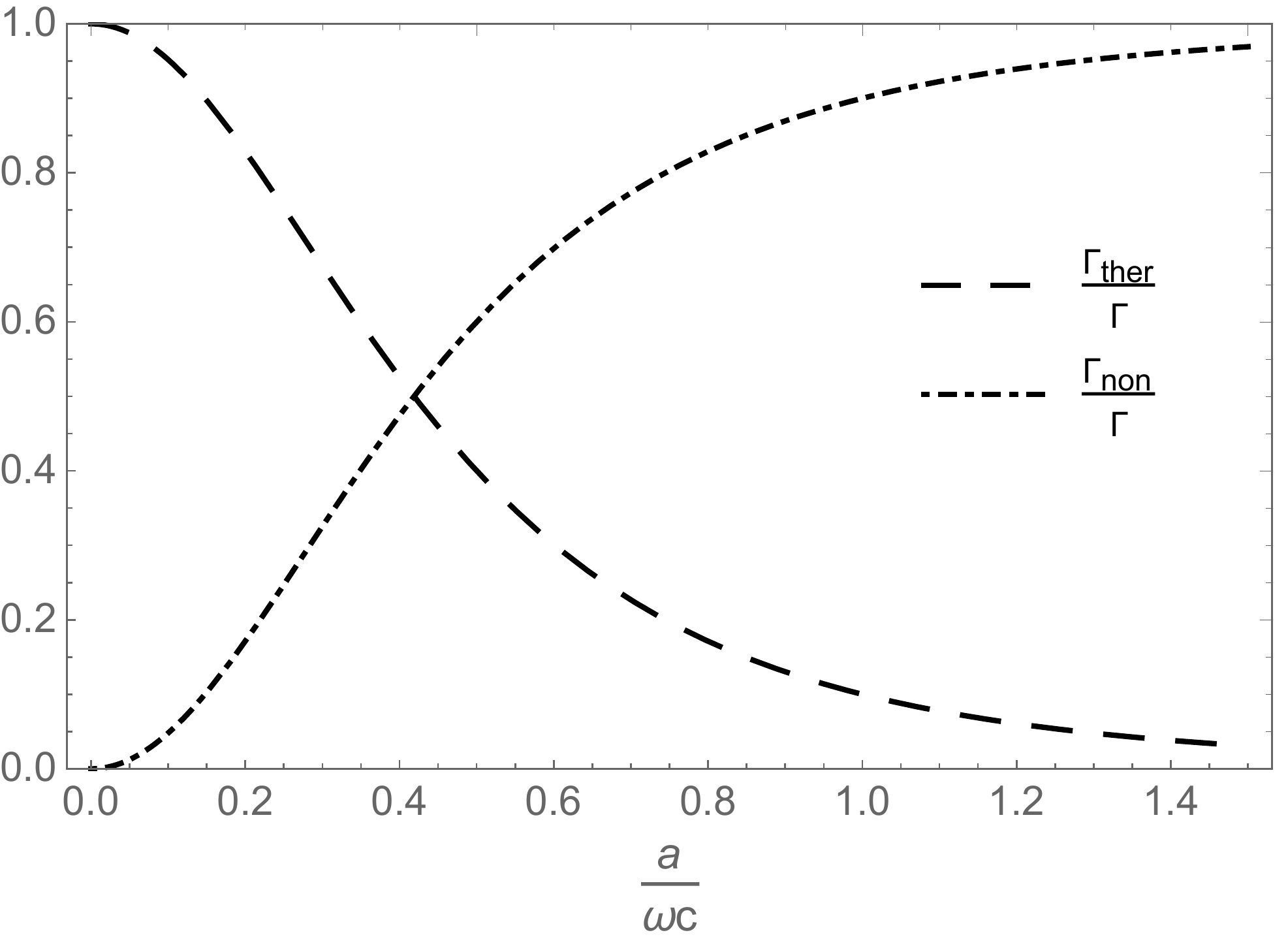}
\caption{\label{1}
The relative weight of the thermal (dashed) and nonthermal (dot-dashed) parts in the total transition rate as a function of acceleration.}
\end{figure}

For accelerated atoms, the transition rate $\Gamma$  depends on acceleration.
When the acceleration $a$ is small compared with $\omega c$,  e.g., $a= 0.1\omega c$, taking the emission rate for inertial atom $\Gamma_{0\downarrow}$ as a reference value, the excitation rate $\Gamma_{\uparrow}$ and emission rate $\Gamma_{\downarrow}$ for accelerated atoms are  $\Gamma_{\uparrow}\thicksim 10^{-28} \Gamma_{0\downarrow}$ and $\Gamma_{\downarrow}\thicksim 1.05 \Gamma_{0\downarrow}$ respectively.
That is, the excitation rate is much smaller than the emission rate, 
which is consistent with the fact that the atom with smaller acceleration is hardly excited from its ground state. 
When the acceleration becomes larger, the excitation rate $\Gamma_{\uparrow}$ becomes more significant. 
For example, when $a= \omega c$, the excitation  and emission rates are  $\Gamma_{\uparrow} \sim 0.019  \Gamma_{0\downarrow}$ and $\Gamma_{\downarrow} \sim 10.019  \Gamma_{0\downarrow}$ respectively. 
As discussed before, there are both thermal and nonthermal parts in the transition rate of a uniformly accelerated atom,  and the relative weights of the thermal and nonthermal parts are
\begin{eqnarray}
&&\frac{\Gamma_{\mathrm{ther}}}{\Gamma}=\frac{1}{1+5(\frac{a}{\omega c})^2+4(\frac{a}{\omega c})^4},\\
&&\frac{\Gamma_{\mathrm{non}}}{\Gamma}=\frac{5(\frac{a}{\omega c})^2+4(\frac{a}{\omega c})^4}{1+5(\frac{a}{\omega c})^2+4(\frac{a}{\omega c})^4},
\end{eqnarray}
respectively, which are the same for both the emission and excitation processes. 
In Fig. \ref{1}, we show how the relative weights of the thermal and nonthermal parts in the total transition rate vary with acceleration. 
It is clear that, for the smaller accelerations, the contributions from thermal terms dominate.
As the acceleration increases, the relative weight of the nonthermal terms increases. When $a$ is larger than $ 0.42\omega c$, contribution from nonthermal terms becomes larger than that of thermal terms. 

\section{summary}


When linear coupling between a gravitationally polarizable atom and the quantum fluctuations of spacetime itself is considered, the rate of change of the atomic energy  is distinctively separated into only two parts, i.e. the contributions of vacuum (thermal) fluctuations and radiation reaction. 
For a uniformly accelerated atom, vacuum fluctuations not only raise the energy of the atom initially in its ground state, but also diminish its energy when the atom is in higher-lying excited states, while radiation reaction always diminishes its energy.
The total rate of change of the energy shows that the perfect balance between the contributions of vacuum (thermal) fluctuations and radiation reaction is disturbed; thus the transition from ground state to higher-lying excited states is possible for both uniformly  accelerated atoms and static ones in a thermal bath of gravitons.
The appearance of power terms in acceleration $a$ in the
mean rate of change of atomic energy  suggests that the equivalence between uniform acceleration and thermal field is lost.

\begin{acknowledgments}

This work was supported in part by the NSFC under Grants No. 11435006, No. 11690034, and No. 11805063, and the Hunan Provincial Innovation Foundation for Postgraduate under Grant No. CX2018B289.

\end{acknowledgments}

\end{document}